\documentclass[journal,twoside]{IEEEtran}

\usepackage{verbatim}
\usepackage{amsmath}
\usepackage{amssymb}
\usepackage[slantedGreek]{mathptmx}
\usepackage{threeparttable}
\usepackage{enumerate}
\usepackage{color}
\usepackage{psfrag}
\usepackage{ifpdf}
\usepackage{multirow}
\usepackage{setspace}
\usepackage{url}
\usepackage{algorithmic}
\usepackage{algorithm}
\usepackage{booktabs}
\usepackage{array}
\usepackage{amsfonts}
\usepackage{graphicx}
\usepackage[numbers,sort&compress]{natbib}
\usepackage[caption=false,font=footnotesize]{subfig}
\usepackage{rotating}
\usepackage{textcomp}
\usepackage{tabularx}
\usepackage{bm}
\usepackage{cases}
\usepackage{color}
\usepackage{soul}
\soulregister\cite7 
\soulregister\citep7
\soulregister\citet7 
\soulregister\ref7
\soulregister\pageref7

\newcolumntype{L}[1]{>{\raggedright\arraybackslash}p{#1}}
\newcolumntype{C}[1]{>{\centering\arraybackslash}p{#1}}
\newcolumntype{R}[1]{>{\raggedleft\arraybackslash}p{#1}}

\begin{document}
%
\title{Deep Low-rank Prior in Dynamic MR Imaging}

%

\author{Ziwen Ke, Wenqi Huang, Jing Cheng,~\IEEEmembership{Student Member,~IEEE}, Zhuoxu Cui, Sen Jia, Haifeng Wang, Xin Liu, Hairong Zheng,~\IEEEmembership{Senior Member,~IEEE}, Leslie Ying,~\IEEEmembership{Senior Member,~IEEE}, Yanjie Zhu
    and~Dong Liang,~\IEEEmembership{Senior Member,~IEEE}

\thanks{This work was supported in part by the National Natural Science Foundation of China (61771463, 81830056, U1805261, 81971611, 61871373, 81729003, 81901736); National Key R$\&$D Program of China (2017YFC0108802 and 2017YFC0112903); Natural Science Foundation of Guangdong Province (2018A0303130132); Shenzhen Key Laboratory of Ultrasound Imaging and Therapy (ZDSYS20180206180631473); Shenzhen Peacock Plan Team Program (KQTD20180413181834876); Innovation and Technology Commission of the government of Hong Kong SAR (MRP/001/18X); Strategic Priority Research Program of Chinese Academy of Sciences (XDB25000000)}
\thanks{Corresponding author: yj.zhu@siat.ac.cn, dong.liang@siat.ac.cn}
\thanks{Ziwen Ke, Wenqi Huang, Zhuoxu Cui and Dong Liang are with Research Center for Medical AI, Shenzhen Institutes of Advanced Technology, Chinese Academy of Sciences, Shenzhen,
	China}
\thanks{Ziwen Ke and Wenqi Huang are also with Shenzhen College of Advanced Technology, University of Chinese Academy
	of Sciences, Shenzhen,
	China}
\thanks{Jing Cheng, Sen Jia, Haifeng Wang, Xin Liu, Hairong Zheng, Yanjie Zhu and Dong Liang are with Paul C. Lauterbur Research Center for Biomedical Imaging, Shenzhen Institutes of Advanced Technology, Chinese Academy of Sciences, Shenzhen, China}
\thanks{Leslie Ying is with Department of Biomedical Engineering and the Department of Electrical Engineering, The State University of New York, Buffalo, NY, USA}
\thanks{Code will be available at https://github.com/Keziwen/SLR-Net}
}

%
%

\markboth{This manuscript has been submitted to a IEEE Journal}
{Ziwen \MakeLowercase{\textit{et al.}}: Deep Low-rank Prior in Dynamic MR Imaging}
%



\maketitle

\begin{abstract}
The deep learning methods have achieved attractive performance in dynamic MR cine imaging. However, all of these methods are only driven by the sparse prior of MR images, while the important low-rank (LR) prior of dynamic MR cine images is not explored, which limits the further improvements on dynamic MR reconstruction. In this paper, a learned singular value thresholding (Learned-SVT) operation is proposed to explore deep low-rank prior in dynamic MR imaging for obtaining improved reconstruction results. In particular, we come up with two novel and distinct schemes to introduce the learnable low-rank prior into deep network architectures in an unrolling manner and a plug-and-play manner respectively. In the unrolling manner, we put forward a model-based unrolling sparse and low-rank network for dynamic MR imaging, dubbed SLR-Net. The SLR-Net is defined over a deep network flow graph, which is unrolled from the iterative procedures in the Iterative Shrinkage-Thresholding Algorithm (ISTA) for optimizing a sparse and low-rank based dynamic MRI model. In the plug-and-play manner, we present a plug-and-play LR network module that can be easily embedded into any other dynamic MR neural networks without changing the network paradigm. Experimental results show that both schemes can further improve the state-of-the-art CS methods, such as k-t SLR, and sparsity-driven deep learning-based methods, such as DC-CNN and CRNN, both qualitatively and quantitatively.
\end{abstract}


%
\IEEEpeerreviewmaketitle

\section{Introduction}
%
%
%
%
\IEEEPARstart{M}{R} cine imaging is of great value in clinical application due to its ability to reveal both spatial anatomical information and dynamic information simultaneously. However, obtaining high spatio-temporal resolution MR cine images is very challenging due to the limited scan time. Therefore, accelerating dynamic MR cine imaging by undersampling k-space has generated great research interest to alleviate these conflicts. 

In the past decade, sparsity prior \cite{1614066, lustig2007sparse, Otazo2010Combination} has been utilized to accelerate MR cine imaging following a path of gradually relaxing constraint. At the earlier stage, the fixed basis like temporal Fourier transform was used to sparsify cine images along the temporal dimension \cite{lustig20006ktsparse, jung2007improved, liang2012k}. Then, some researchers proposed that the basis is not restricted to the fixed ones, and they used dictionary learning to learn an adaptive basis solely from acquired data \cite{caballero2014dictionary, wang2013compressed, nakarmi2015dynamic}. Very recently, the sparsity constraint is relaxed further by learning sparsifying transform from large data sets using an unrolling-based deep learning technique \cite{schlemper2018deep, qin2018convolutional, shan2019dimension}. 
With the relaxed sparse constraint, the reconstruction performance becomes more and more superior given enough training data. Moreover, deep learning based methods avoid the issues of relatively long reconstruction time and the empirical selection of the regularization parameters \cite{Dong2020Deep}. 

As an extension of sparsity, low-rank prior has been applied to dynamic MRI \cite{Liang2007Spatiotemporal, Justin2010Spatiotemporal, lingala2011accelerated, zhao2012image, otazo2015low}, and the low-rank based dynamic MRI methods take on a similar path with what the sparse prior undergoes. First, by considering each temporal frame as a column of a space-time matrix, the low-rank property from the spatio-temporal correlations of dynamic MR images was pursued by minimizing the number of non-zero singular values of this matrix. By suppressing the singular vectors of the low-rank matrix that correspond to artifacts and noise, these methods could obtain improved reconstruction. Recently, Manifold-learning based dynamic MRI methods were developed \cite{Nakarmi2017A, shetty2019bi, poddar2015dynamic}, where the low-rank property was exploited in a low-dimension manifold which was learned from acquired k-space only. Although these methods have shown that dynamic MR images have a strong low-rank prior that can be used to improve the reconstruction, such a prior has not been utilized in any deep learning based methods.

In this paper, we exploit the deep low-rank prior in dynamic MR imaging in an unrolling manner and a plug-and-play (PnP) manner, respectively. In the unrolling scheme, we formulate the sparse and low-rank priors as regularized terms in an optimization problem and unroll a dynamic MR model to a deep neural network and. Importantly, a learned singular value thresholding (Learned-SVT) operation is proposed to learn the low-rank prior using deep learning techniques. The comparisons with the traditional k-t SLR \cite{lingala2011accelerated} and a sparsity-driven deep learning based method ISTA-Net \cite{Zhang_2018_CVPR} on in vivo data sets show that the proposed SLR-Net method achieves superior reconstruction performance. However, in this scheme, the optimization algorithm is highly customized, and cannot apply low-rankness as a prior to other general inverse problems. To benefit general dynamic imaging problems, we further come up with a plug-and-play scheme, which starts with the same model but uses a first-order Taylor approximation for both the data fidelity term and the regularization term. This plug-and-play LR network module can be easily embedded into any sparsity-driven deep learning models such as DC-CNN \cite{schlemper2018deep} and CRNN \cite{qin2018convolutional} without changing the network paradigm. The addition of the low-rank prior consistently improves deep learning based dynamic MR reconstruction upone the methods without such a prior. To the best of our knowledge, this is the first time that a deep low-rank prior is applied in dynamic MR imaging. 

The rest of this paper is organized as follows. Section II provides the background and introduces the proposed methods. Section III summarizes experimental details and the results to demonstrate the effectiveness of the proposed method, while the discussion and conclusions are presented in Section IV and Section V, respectively.

\section{Methodology}
\subsection{Background: Dynamic MR Imaging with Learned Sparse Prior}
The goal of dynamic MRI is to estimate a sequence of complex-valued MR images $\bm{x}\in\mathbb{C}^{N_xN_yN_t}$ from the undersampled k-space measurements $\bm{y}\in \mathbb{C}^{N_xN_yN_t}$.  This problem is commonly formulated as the following optimization problem with sparse prior:
\begin{equation}
\label{eq_1}
\bm{x}^* = \arg\min_{\bm{x}}\  \frac{1}{2}||\bm{Ax}-\bm{y}||_2^2+\lambda ||\bm{D}\bm{x}||_1
\end{equation}
Here $\bm{A}=\bm{PF}$ is the encoding operator, $\bm{F}$ is a Fourier transform and $\bm{P}$ is a under-sampling matrix. The first term is the data fidelity, which ensures that the k-space of reconstruction is consistent with the actual measurements. The second term is often referred to as the prior regularization. $\lambda$ is a regularization parameter. In CS-based methods, $\bm{D}$ is the sparsifying transform. 
An auxiliary variable $\bm{z}$, which is constrained to be equal to $\bm{D}\bm{x}$, is usually introduced to decouple the fidelity term and the regularization
term as follows:
\begin{equation}
\label{eq_2}
\arg\min_{\bm{x}, \bm{z}}\ \frac{1}{2}||\bm{Ax}-\bm{y}||_2^2+\lambda ||\bm{z}||_1+\alpha||\bm{D}\bm{x}-\bm{z}||_2^2
\end{equation}
where $\alpha$ is a penalty parameter. This variable splitting technique can be solved iteratively by applying an alternating minimization algorithm:
\begin{equation}
\label{eq_3}
\begin{aligned}
\bm{z}^{n+1} & =\arg\min_{\bm{z}}\ \lambda ||\bm{z}||_1+\alpha||\bm{D}\bm{x}^{n+1}-\bm{z}||_2^2 \\
\bm{x}^{n+1} & = \arg\min_{\bm{x}}\ \frac{1}{2}||\bm{Ax}-\bm{y}||_2^2+\alpha||\bm{D}\bm{x}-\bm{z}^n||_2^2 \\
\end{aligned}
\end{equation}
The second equation of Eq.\ref{eq_3} is often performed by a backfill operation on k-space, which is known as a data consistency (DC) step. For the k-space coefficients that are initially unknown, the reconstructed values are used. For the coefficients that have already been sampled, the predicted k-space is updated with the combination of the actual sampled k-space and the predicted k-space. The formula and more details about DC can be found in \cite{shan2019dimension}. The optimization process of Eq.\ref{eq_3} can be summarized in the following paradigm:
\begin{equation}
\label{eq_4}
\bm{x}^0\rightarrow\bm{z}^1\stackrel{DC}{\longrightarrow}\bm{x}^1\rightarrow\cdots\rightarrow\bm{z}^N\stackrel{DC}{\longrightarrow}\bm{x}^N
\end{equation}
where $N$ is the total number of iterations.

Existing deep learning methods for dynamic MR imaging, such as DC-CNN \cite{schlemper2018deep}, CRNN \cite{qin2018convolutional}, and DIMENSION \cite{shan2019dimension}, unroll the Eq.\ref{eq_3} into specific neural networks in a cascaded \cite{schlemper2018deep, shan2019dimension} or recurrent \cite{qin2018convolutional} manner. Their network topologies also satisfy the above paradigm. Although there are a number of related works on arXiv, here we only focus on the peer-reviewed papers.

\subsection{The Proposed Methods: Dynamic MR Imaging with Learned Both Low Rank and Sparse Prior}
In recent years, a large number of works \cite{Liang2007Spatiotemporal, Justin2010Spatiotemporal, lingala2011accelerated, otazo2015low, Nakarmi2017A} based on low-rank matrix completion have proved that dynamic MR images have a strong low-rank prior, which can be used to improve the reconstruction. 

Regularized matrix recovery or low-rank matrix completion, exploits the compact signal representation in the Karhunen Louve Transform (KLT) domain \cite{Liang2007Spatiotemporal, Justin2010Spatiotemporal, Brinegar2008Real, Pedersen2009k, lingala2011accelerated}. Different from Eq.\ref{eq_1}, which only uses the sparse prior of dynamic images, an additional low-rank prior can be introduced into the dynamic MR model to obtain the following optimization equation:
\begin{equation}
\label{eq_5}
\bm{x}^* = \arg\min_{\bm{x}} \ \frac{1}{2}||\bm{Ax}-\bm{y}||_2^2+\lambda_1 ||\bm{D}\bm{x}||_1 + \lambda_2||\bm{x}||_*
\end{equation}
where $||\bm{x}||_*$ is the nuclear norm, which is the sum of the singular values of $\bm{x}$ ($||\bm{x}||_*=\sum_{i}(\bm{\Sigma}_i,i), \bm{x}=\bm{U}\bm{\Sigma}\bm{V}^*$). The regularized matrix recovery using the nuclear norm minimization has been rigorously studied \cite{Cand2008Exact, Cai2010A, Lee2009ADMiRA}. In our work, we will use the ISTA solver \cite{Beck2009A} to solve Eq.\ref{eq_5}.

By introducing auxiliary variable $\bm{t}$, the fidelity term with the sparse regularization term and the low-rank regularization term can be decoupled:
\begin{equation}
\label{eq_6}
\arg\min_{\bm{x}, \bm{t}}\ \frac{1}{2}||\bm{Ax}-\bm{y}||_2^2+\lambda_1 ||\bm{D}\bm{x}||_1+\lambda_2 ||\bm{t}||_*\ \ \ s.t.\ \bm{t}=\bm{x}
\end{equation}

In this paper, we come up with two schemes to solve Eq.\ref{eq_6} by introducing a learnable low-rank prior to the deep network architectures, as detailed below.

\subsubsection{The First Scheme: The Proposed Unrolling Manner}

By applying an augmented Lagrangian function, Eq.\ref{eq_6} is turned into an unconstrained optimization problem:
\begin{equation}
\label{eq_7}
\begin{aligned} 
\mathcal{L}(\bm{x}, \bm{t}, \bm{\alpha})&=  \frac{1}{2}||\bm{Ax}-\bm{y}||_2^2+\lambda_1 ||\bm{D}\bm{x}||_1+\lambda_2 ||\bm{t}||_* \\
&\ \ \  - \langle\bm{\alpha}, \bm{t}-\bm{x}\rangle + \frac{\rho}{2}||\bm{t}-\bm{x}||_2^2 
\end{aligned} 
\end{equation}
where $\bm{\alpha}$ is a Lagrangian multiplier and $\rho$ is a penalty parameter. A proximal point algorithm (PPA) \cite{Rockafellar1976Monotone} is applied to express the subproblems as
\begin{equation}
\label{eq_8}
\begin{cases} 
\arg\min\limits_{\bm{x}}\  \frac{1}{2}||\bm{Ax}-\bm{y}||_2^2+ \frac{\rho}{2}||\bm{x}+\bm{\beta}-\bm{t}||_2^2 +\lambda_1||\bm{D}\bm{x}||_1 \\

\arg\min\limits_{\bm{t}}\ \frac{\rho}{2}||\bm{x}+\bm{\beta}-\bm{t}||_2^2+\lambda_2||\bm{t}||_*\\ 

\bm{\beta}\leftarrow\bm{\beta}+\tilde{\eta}_1(\bm{x}-\bm{t})
\end{cases} 
\end{equation}
where $\bm{\beta}=\frac{\bm{\alpha}}{\rho}$ is a scaled Lagrangian multiplier and $\tilde{\eta}_1$ is an update rate. The subproblem of $\bm{x}$ is a general $l_1$ norm CS reconstruction model. The iterative shrinkage-thresholding algorithm (ISTA) \cite{Beck2009A} is
a popular first-order proximal method, which is well suited for solving this subproblem. Specifically, ISTA solves the subproblem of $\bm{x}$ by iterating between the following update steps: 
\begin{equation}
\label{eq_9}
\begin{cases}
\bm{r}^n=& \bm{x}^{n-1}-\tilde{\eta}_2(\bm{A}^T(\bm{A}\bm{x}^{n-1}-\bm{y})+\rho(\bm{x}^{n-1}+\bm{\beta}^{n-1}-\bm{t}^{n-1}))\\
\bm{x}^n=&\arg\min\limits_{\bm{x}}||\bm{x}-\bm{r}^n||_2^2+\lambda_1||\bm{D}\bm{x}||_1
\end{cases}
\end{equation}
where $\tilde{\eta}_2$ is an update rate. Substituting Eq.\ref{eq_9} into Eq.\ref{eq_8}, we get
\begin{equation}
\label{eq_10}
\begin{cases} 
\bm{r}^n=& \bm{x}^{n-1}-\tilde{\eta}_2(\bm{A}^T(\bm{A}\bm{x}^{n-1}-\bm{y})+\rho(\bm{x}^{n-1}+\bm{\beta}^{n-1}-\bm{t}^{n-1}))\\
\bm{x}^n=&\arg\min\limits_{\bm{x}}||\bm{x}-\bm{r}^n||_2^2+\lambda_1||\bm{D}\bm{x}||_1\\
\bm{t}^n=&\arg\min\limits_{\bm{t}}\ \frac{\rho}{2}||\bm{x}^n+\bm{\beta}^{n-1}-\bm{t}||_2^2+\lambda_2||\bm{t}||_*\\ 
\bm{\beta}^n=&\bm{\beta}^{n-1}+\tilde{\eta}_1(\bm{x}^n-\bm{t}^n)
\end{cases} 
\end{equation}
For the subproblem of $\bm{x}^n$, how to solve $\bm{x}^n$ effectively and efficiently is critical. Optimization methods such as ADMM \cite{Afonso2011An} and AMP \cite{Metzler2016From} provide effective methods to solve $\bm{x}^n$. When the sparse transform $\bm{D}$ is orthogonal (for example $\bm{D}$ is wavelet transform), we can obtain $\bm{x}$ by simple threshold operation: $\bm{x}^n=\bm{D}^T\text{soft}(\bm{D}\bm{r}^n, \lambda_1)$. However, it remains non-trivial to solve $\bm{x}^n$ for a non-orthogonal transform $\bm{D}$. For simplicity of calculation, this paper focuses on the case where $\bm{D}$ is orthogonal. For the subproblem of $\bm{t}^n$, an iterative singular value thresholding (IST) scheme \cite{Cai2010A} is used in the nuclear norm minimization and we have
\begin{equation}
\label{eq_11}
 \bm{t}^n=IST(\bm{x}^{n})=\sum\limits_{i=0}^{\min(m,n)}(\Sigma_i-\lambda_2\Sigma_i^{p-1}/\rho)_+\bm{u}_i\bm{v}_i^*
\end{equation}
where $\bm{u}_i, \bm{v}_i$ and $\Sigma_i$ are the singular vectors and values of $\bm{x}^n$. The thresholding function $(\sigma)_+$ is defined as 
\begin{equation}
\label{eq_12}
(\sigma)_+=
\begin{cases}
\sigma, \ \ \ \text{if}\ \ \sigma\ge 0\\
0, \ \ \ \text{else}
\end{cases}
\end{equation}

Finally, the sparse and low-rank MR model obtains the following iterative procedures:
\begin{equation}
\label{eq_13}
\begin{cases}
\bm{R}^n:\ &\bm{r}^n=\bm{x}^{n-1}-\tilde{\eta}_2(\bm{A}^T(\bm{A}\bm{x}^{n-1}-\bm{y})+\rho(\bm{x}^{n-1}+\bm{\beta}^{n-1}-\bm{t}^{n-1}))\\
\bm{X}^n:\ 
&\bm{x}^n=\bm{D}^T\text{soft}(\bm{D}\bm{r}^n, \lambda_1)\\
\bm{T}^n:\ 
&\bm{t}^n=IST(\bm{x}^{n})=\sum\limits_{i=0}^{\min(m,n)}(\Sigma_i-\lambda_2\Sigma_i^{p-1}/\rho)_+\bm{u}_i\bm{v}_i^*\\
\bm{M}^n:\ 
&\bm{\beta}^n=\bm{\beta}^{n-1}+\tilde{\eta}_1(\bm{x}^n-\bm{t}^n)
\end{cases} 
\end{equation}

In traditional CS-MRI, an optimized reconstruction result $\bm{x}^*$ can be obtained by iteratively solving Eq.\ref{eq_13}. However, both the hyper-parameters $\{\lambda_1, \lambda_2, \rho, \tilde{\eta}_1, \tilde{\eta}_2\}$ and the sparse transform $\bm{D}$ need to be selected empirically, which is tedious and uncertain. What's worse, the iterative solution often takes a long time, which limits its clinical application.

\begin{figure}[htbp]
	\centerline{\includegraphics[width=1.0\linewidth]{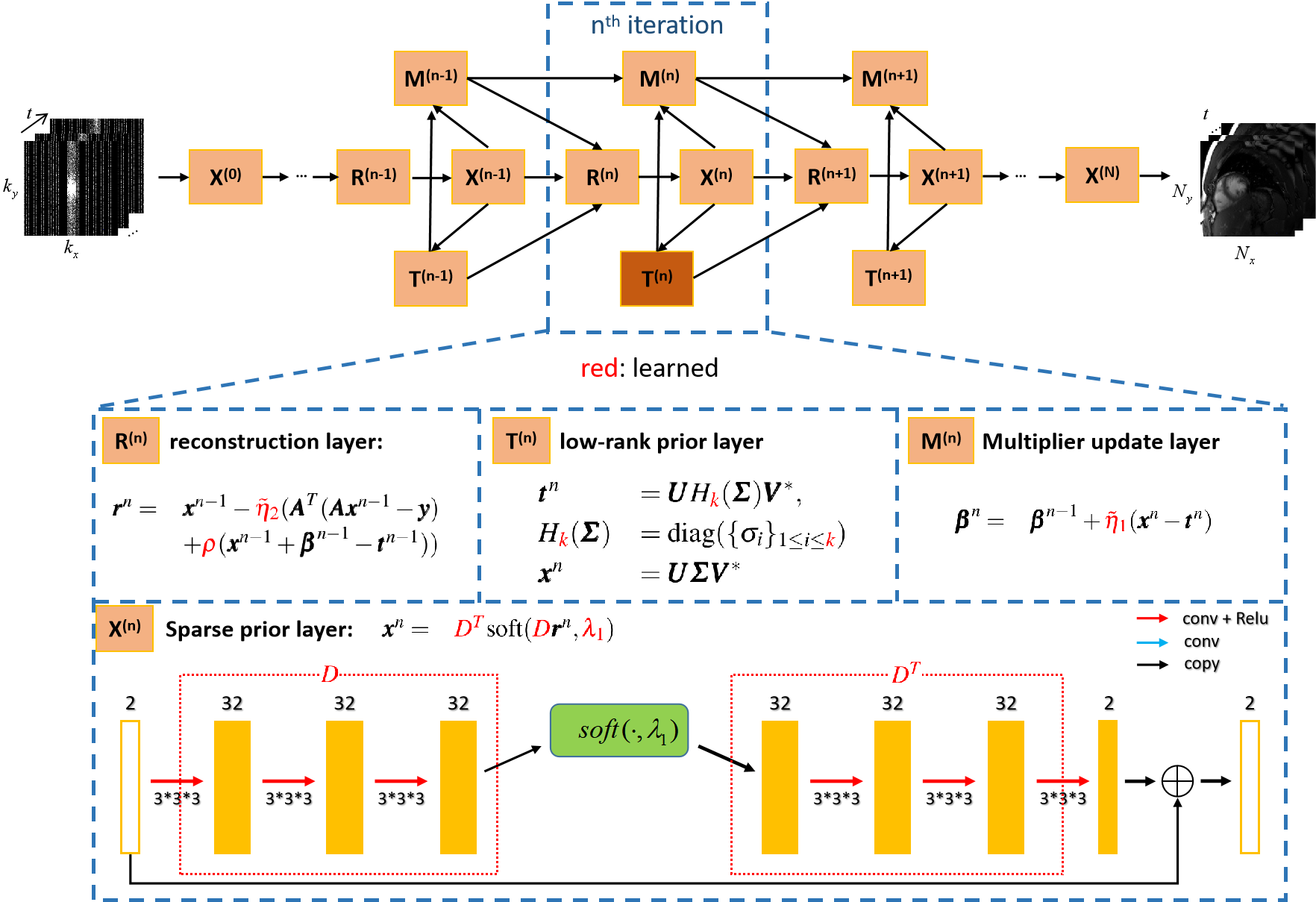}}
	\caption{The proposed sparse and low-rank network (SLR-Net) for dynamic MRI. The SLR-Net is defined over the iterative procedures of Eq.\ref{eq_13}. The four procedures in Eq.\ref{eq_13} correspond to the four modules in SLR-Net, which are named as reconstruction layer $\bm{R}^n$, sparse prior layer $\bm{X}^n$, low-rank prior layer $\bm{T}^n$, and multiplier update layer $\bm{M}^n$ respectively. SLR-Net keeps the same arithmetic structures, but its hyper-parameters and sparse transform are learnable. \label{SLR-Net}}
\end{figure}

To address the above-mentioned issues, we propose a deep sparse and low-rank network, dubbed as SLR-Net. It unrolls the above iterations into a deep neural network. Specifically, the four procedures in Eq.\ref{eq_13} correspond to the four modules in SLR-Net as shown in Fig.\ref{SLR-Net}, which are named as reconstruction layer $\bm{R}^n$, sparse prior layer $\bm{X}^n$, low-rank prior layer $\bm{T}^n$, and multiplier update layer $\bm{M}^n$. SLR-Net keeps the same arithmetic structures, but its hyper-parameters $\{\lambda_1, \lambda_2, \rho, \tilde{\eta}_1, \tilde{\eta}_2\}$ and the sparse transform $\bm{D}$ are learnable. We will discuss the four modules in detail next.
\begin{itemize}
	\item{\textbf{Reconstruction layer} $\bm{R}^n$:}
	The reconstruction result of the current iteration or layer $\bm{r}^n$ can be obtained under the given $\{\bm{x}^{n-1}, \bm{\beta}^{n-1}, \bm{t}^{n-1}\}$ according to Eq.\ref{eq_13}. The hyper-parameter $\rho$ and $\tilde{\eta}_2$ are set as learnable network parameters, which initialized to zero and 0.1, respectively. When $n=1$, $\bm{x}^0$ is the zero-filling image, and $\{\bm{\beta}^{0}, \bm{t}^{0}\}$ are initialized to zeros. 
	\item {\textbf{Sparse prior layer} $\bm{X}^n$:}
	This layer explores the sparse prior of the current reconstruction $\bm{r}^n$. $\bm{x}^n$ can be obtained according to Eq.\ref{eq_13}. Unlike traditional CS-MRI, where the sparse transform $\bm{D}$ is designed empirically, the SLR-Net can learn a general sparse transform $\bm{D}$ using convolutional neural networks. We don't strictly require $\bm{D}$ and $\bm{D}^T$ to be transpose to each other. Instead, they are learned by different networks $\{\bm{\tilde{D}}_{1}, \bm{\tilde{D}}_{2}\}$ to increase network capacity. Three convolutional layers are used to learn the transforms.
	\item {\textbf{Low-rank prior layer} $\bm{T}^n$:} 
	To solve the low-rank constrained subproblem, IST is used in nuclear norm minimization with a soft threshold scheme. In this paper, we proposed a learnable low-rank prior for the first time:
	\begin{equation}
	\label{eq_14}
	\begin{cases}
	\bm{t}^n&=\bm{U}\mathit{H}_k(\bm{\Sigma})\bm{V}^*, \\
	\mathit{H}_k(\bm{\Sigma}) &= \text{diag}(\{\sigma_i\}_{1\leq i\leq k})\\
	\bm{x}^{n}&=\bm{U}\bm{\Sigma}\bm{V}^*
	\end{cases} 
	\end{equation}
	where $\bm{x}^{n}=\bm{U}\bm{\Sigma}\bm{V}^*$ is the singular value decomposition of $\bm{x}^{n}$. $\mathit{H}_k$ is a learned singular value thresholding (SVT) operation. Specifically, the singular values of top $k$ are completely retained, and the remaining singular values are removed for suppression. The hyper-parameter $k$ is set as a learnable network parameter, which is initialized to one.
	\item {\textbf{Multiplier update layer} $\bm{M}^n$:}
	This layer is used to update the Lagrange multiplier $\bm{\beta}$. $\tilde{\eta}_1$ is a learned parameter.	 
\end{itemize}


\subsubsection{The Second Scheme: The Proposed Plug-and-play Manner}
In the first scheme, the low-rank prior of dynamic data has been introduced by unrolling a sparse and low-rank based optimization algorithm. However, the optimization algorithm is highly customized, and currently, no simple published turn-key deep learning methods exist to apply low-rankness as a prior to other general inverse problems. In \cite{6737048}, plug-and-play priors are integrated with the forward model of imaging systems. Inspired by this idea, we provide another scheme to exploit the low-rank prior in a network via a plug-and-play LR network module for a general inverse problem.

Here, we propose to use the ISTA algorithm for solving problem (Eq.\ref{eq_6}), where the combination of the sparsity promoting regularizer $\lambda_1||\bm{Dx}||_1$ and the low rank regularizer $\lambda_2||\bm{x}||_*$ is considered. 
Approximating the objective of (Eq.\ref{eq_6}) by a first-order Taylor approximation for the data fidelity $\|\bm{Ax}^n-\bm{y}\|$ at $\bm{x}^{n-1}$ and for the regularizer $\lambda_1\|\bm{Dx}\|_1$ at $\bm{x}^n$, at the $n$-th iteration, we derive an approximated objective function as follows:
\begin{equation}
\label{ite}
\begin{aligned}
J(\bm{t}):=& \langle \bm{A}^T(\bm{Ax}^{n-1}-\bm{y}),\bm{t}-\bm{x}^{n-1}\rangle+\langle\lambda_1 g^n,\bm{t}-\bm{x}^n \rangle\\&+ \frac{1}{2\widetilde{\eta}}\|\bm{t}-\bm{x}^{n-1}\|^2+\lambda_2\|\bm{t}\|_*
\end{aligned}
\end{equation}
where $g^n\in \partial\|\bm{Dx}^n\|$, and $\bm{x}^n$ is a point updated by an ISTA step about regularizer $\lambda_1\|\bm{Dx}\|_1$ further, in particular, which can be formulated as follows: 
\begin{equation}\label{ista}\begin{aligned}
\bm{x}^n=&\arg\min_{\bm{x}}\langle \bm{A}^T(\bm{Ax}^{n-1}-\bm{y}),\bm{x}-\bm{x}^{n-1}\rangle+\frac{1}{2\widetilde{\eta}}\|\bm{x}-\bm{x}^{n-1}\|^2\\&+\lambda_1\|\bm{Dx}\|_1\\
=&\arg\min_{\bm{x}}\frac{1}{2\widetilde{\eta}}\|\bm{x}-(\bm{x}^{n-1}-\widetilde{\eta}\bm{A}^T(\bm{Ax}^{n-1}-\bm{y})) \|^2\\&+\lambda_1\|\bm{Dx}\|_1
\end{aligned}\end{equation}
To update the variables according to the rule of ISTA, i.e., $\bm{t}^n=\arg\min_{\bm{t}}J(\bm{t})$, we have
\begin{equation}
\label{ite2}
\begin{aligned}
\bm{t}^n=&\arg\min_{\bm{t}}\frac{1}{2\widetilde{\eta}}\|\bm{t}-(\bm{x}^{n-1}-\widetilde{\eta}\bm{A}^T(\bm{Ax}^{n-1}-\bm{y})-\widetilde{\eta}\lambda_1g^n) \|^2\\&+\lambda_2\|\bm{t}\|_* 
\end{aligned}
\end{equation}
Then, we propose to decompose the update rule (\ref{ite2}) into a concise 2-steps update: 
Firstly, based on the first-order optimization condition of (\ref{ista}), we have
\begin{equation}\label{prox1}\bm{x}^n= \bm{x}^{n-1}- \widetilde{\eta}\bm{A}^T(\bm{Ax}^{n-1}-\bm{y})-\widetilde{\eta}\lambda_1g^n.\end{equation}
which can be reformulated as  
\begin{equation}\label{prox3}\begin{aligned}
\bm{r}^n=&\bm{x}^{n-1}- \widetilde{\eta}\bm{A}^T(\bm{Ax}^{n-1}-\bm{y})\\
\bm{x}^n=&\bm{D}^T\text{soft}(\bm{Dr}^n,\lambda_1\widetilde{\eta})
\end{aligned}\end{equation}
Secondly, substituting (\ref{prox1}) into (\ref{ite}), we have
\begin{equation}\label{prox2} \bm{t}^n=\bm{U}H_k(\bm{\Sigma})\bm{V}^T\end{equation}
Based on the definition of nuclear norm proximal operator. With (Eq.\ref{prox2}) and (Eq.\ref{prox3}) together, the following formula is derived:
\begin{equation}
\label{ista-lr}
\begin{cases}
\bm{R}^n:\ &\bm{r}^n=\bm{x}^{n-1}-\tilde{\eta}(\bm{A}^T(\bm{A}\bm{x}^{n-1}-\bm{y}))\\
\bm{X}^n:\ 
&\bm{x}^n=\bm{D}^T\text{soft}(\bm{D}\bm{r}^n, \lambda)\\
\bm{T}^n:\ 
&\bm{t}^n=\bm{U}\mathit{H}_k(\bm{\Sigma})\bm{V}^*\\
\end{cases} 
\end{equation}

\begin{figure}[htbp]
	\centerline{\includegraphics[width=1.0\linewidth]{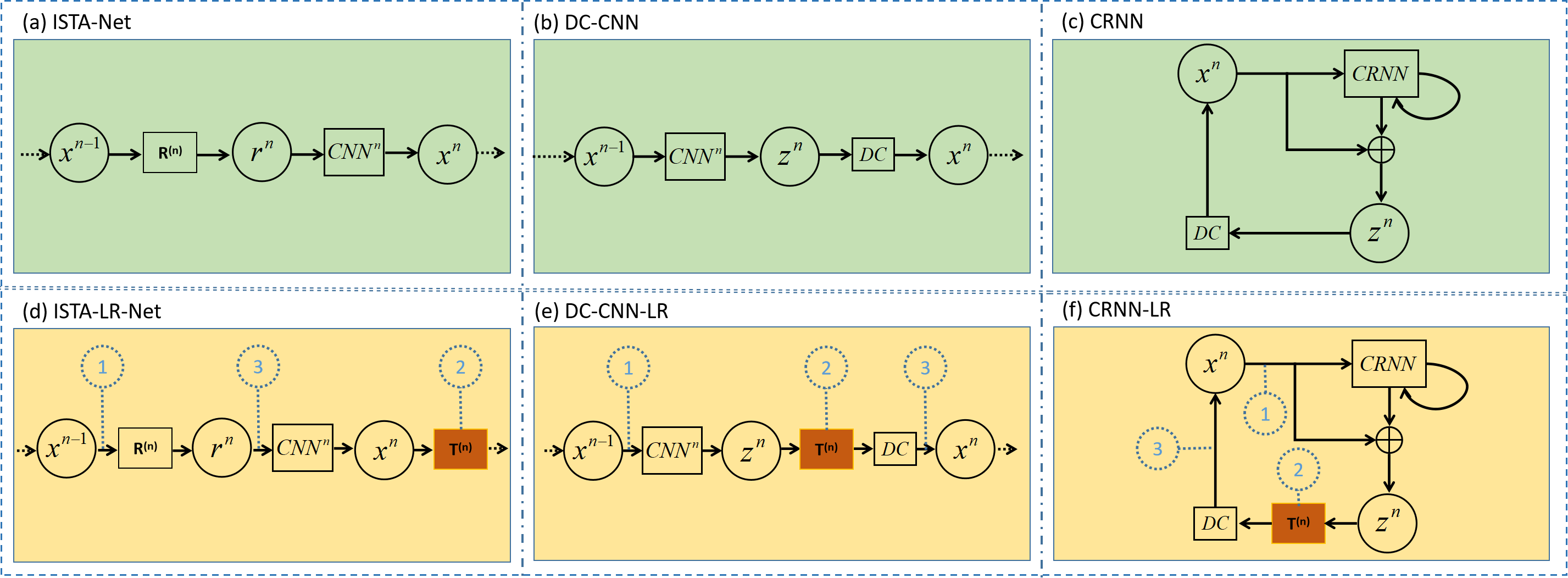}}
	\caption{The proposed plug-and-play LR network module. (a) The original ISTA-Net. (b) The original DC-CNN. (c) The original CRNN. (d) ISTA-LR-Net by embedding the LR network module into the original ISTA-Net. (e) DC-CNN-LR by embedding the LR network module into the original DC-CNN.(f) CRNN-LR by embedding the LR network module into the original CRNN. The numbers in the dotted box represent the locations where the LR module can be embedded. \label{LR-Module}}
\end{figure}

Unrolling the above iterations (Eq.\ref{ista-lr}) into a deep neural network, we develop a new network, dubbed as ISTA-LR-Net shown in Fig.\ref{LR-Module} (d). The iterative procedure in (Eq.\ref{prox3}) is the unrolling processes of ISTA-Net \cite{Zhang_2018_CVPR}, which is an unrolled network developed by applying the ISTA algorithm to the sparse-driven optimization problem in Eq.\ref{eq_1}. Its network topology can be found in Fig.\ref{LR-Module} (a). Intuitively, ISTA-LR-Net can be seen as embedding a low-rank module into the network topology of ISTA-LR-Net in location L2 (Fig.\ref{LR-Module} (d), L2). If the low-rank module, $\bm{T}^{n}$, is plugged in other locations (L1, L3) (please see Fig.\ref{LR-Module} (d) for detail), we can make a similar derivation for ISTA-LR-Net only by changing the point of the Taylor expansion (Eq.\ref{ite2}).

We can also treat the low-rank module as a plug-and-play block, which can be easily embedded into other deep learning models without changing the network paradigm. This paper provides three instances of how to embed the proposed learned LR module as shown in Fig.\ref{LR-Module}. Fig.\ref{LR-Module} (a, b, c) represent the original ISTA-Net \cite{Zhang_2018_CVPR},  DC-CNN \cite{schlemper2018deep}, and CRNN \cite{qin2018convolutional}, respectively. Fig.\ref{LR-Module} (d, e, f) represent their respective models embedded with the plug-and-play LR module, which dubbed as ISTA-LR-Net, DC-CNN-LR, and CRNN-LR, respectively.  

The learned LR module $\bm{T}^n$ performs SVT operations on the input signal as shown in Eq.\ref{eq_14}. If the LR module is embedded before the CNN network (L1), the low-rank prior are loaded into the dynamic signal, and the network will learn the sparse and low-rank features of the signal. If the LR module is embedded after the CNN network (L2, L3), the LR module is used for low-rank correction of predicted results to ensure its low-rank prior, just like the DC module for k-space correction of the predicted results. The embedded LR module implies that the model favors the low-rank solution.

\section{Experimental Results}
\subsection{Setup}
\subsubsection{Data acquisition}
The fully sampled cardiac cine data were collected from 30 healthy volunteers on a 3T scanner (MAGNETOM Trio, Siemens Healthcare, Erlangen, Germany) with a 20-channel receiver coil array. All in vivo experiments were approved by the Institutional Review Board (IRB) of Shenzhen Institutes of Advanced Technology, and informed consent was obtained from each volunteer. For each subject, 10 to 13 short-axis slices were imaged with the retrospective electrocardiogram (ECG)-gated segmented bSSFP sequence during breath-hold. Total 386 slices were collected. The following sequence parameters were used: FOV = $330\times330$ mm, acquisition matrix = $256\times256$, slice thickness = 6 mm, TR/TE = 3.0 ms/1.5 ms. The acquired temporal resolution was 40.0 ms and reconstructed to produce 25 phases to cover the entire cardiac cycle. The raw multi-coil data of each frame was combined by the adaptive coil combine method \cite{Walsh2000Adaptive} to produce a single-channel complex-valued image. We randomly selected 25 volunteers for training and the rest for testing. Deep learning typically requires a large amount of data for training \cite{lecun2015deep}. Therefore, we applied data augmentation using rigid transformation-shearing. We sheared the dynamic images along the x, y, and t directions. The sheared size was $192\times192\times16$ ($x\times y\times t$), and the stride along the three directions is 15, 15, and 7, respectively. Finally, we obtained 1548 2D-t cardiac MR data of size $192\times192\times16$ ($x\times y\times t$) for training and 45 data for testing.

Retrospective undersampling was performed to generate input/output pairs for network training. For each frame, we fully sampled frequency-encodes (along $k_x$) and randomly undersampled phase-encodes (along $k_y$) according to a zero-mean Gaussian variable density function \cite{jung2007improved} as shown in Fig.3. Wherein 4 central phase-encodes were ensured to be sampled.

\begin{figure}[htbp]
	\centerline{\includegraphics[width=1\linewidth]{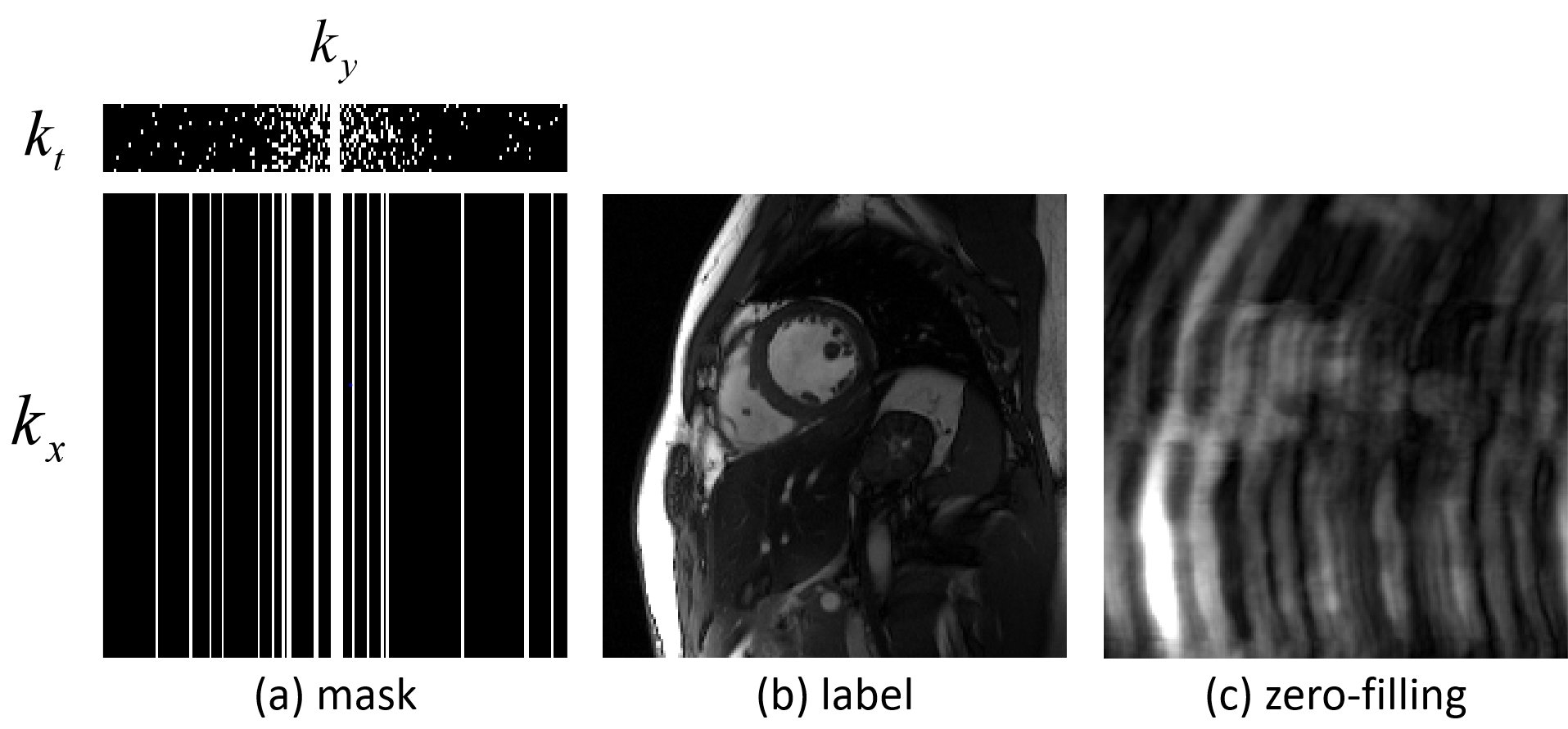}}
	\caption{The gaussian variable density undersampling mask used in this work at 8-fold. (a) mask. (b) label. (c) zero-filling image. \label{mask}}
\end{figure}

\subsubsection{Model configuration}
To demonstrate the effectiveness of the unrolling manner in dynamic MR cine imaging, we compared it with classical k-t SLR \cite{lingala2011accelerated}. The k-t SLR method selected its single-channel versions. For fair comparisons, we adjusted the parameters of the k-t SLR method to its best performance. We also compared it with ISTA-Net, which only explores sparse prior of dynamic MR images, to demonstrate the positive effects of the deep low-rank prior in SLR-Net.

To demonstrate the effectiveness of the plug-and-play LR manner, we embedded it in the state-of-the-art CNN-based methods, DC-CNN \cite{schlemper2018deep} and CRNN \cite{qin2018convolutional} methods, as shown in Fig.\ref{LR-Module}. They were executed according to the source code provided by the authors. For the DC-CNN method, we focused on a D5C5 model, which
works pretty well for this method.

All the CNN-based methods keep the same hyper-parameters. For network training, we divided each data into two channels, where the channels stored real and imaginary parts of the data. So the inputs of the network are undersampled k-space $\mathbb{C}^{2N_xN_yN_t}$ and the outputs are reconstruction images $\mathbb{C}^{2N_xN_yN_t}$. SLR-Net has 8 iterative steps, that is, $N=8$. Each convolutional layer has 32 convolution kernels, and the size of each convolution kernel is $3 \times 3 \times 3$. He initialization \cite{he2015delving} was used to initialize the network weights. Rectifier Linear Units (ReLU) \cite{glorot2011deep} were selected as the nonlinear activation functions. The mini-batch size was 4. The exponential decay learning rate \cite{zeiler2012adadelta} was used in all CNN-based experiments, and the initial learning rate was set to 0.001 with a decay of 0.95. All the models were trained by the Adam optimizer \cite{kingma2014adam} with parameters $\beta_1=0.9$, $\beta_2=0.999$, and $\epsilon=10^{-8}$.

The models were implemented on an Ubuntu 16.04 LTS (64-bit) operating system equipped with an Intel Xeon Gold 5120 Processor Central Processing Unit (CPU) and Tesla V100 Graphics Processing Unit (GPU, 32GB memory) in the open framework Tensorflow \cite{abadi2016tensorflow} with CUDA and CUDNN support. The network training took approximately 18 hours within 50 epochs. 

\subsubsection{Performance evaluation}
For a quantitative evaluation, mean square error (MSE), peak signal to noise ratio (PSNR) and structural similarity index (SSIM) \cite{wang2004image} were measured as follows:
\begin{equation}
\label{eq_15}
\mathrm{MSE}=
||Ref-Rec||^2_2
\end{equation}
\begin{equation}
\label{eq_16}
\mathrm{PSNR}
= 20\log_{10}\frac{\max(Ref)\sqrt{N}}{||Ref-Rec||_2}
\end{equation}
\begin{equation}
\label{eq_17}
\mathrm{SSIM}
= \boldsymbol{l}(Ref, Rec)\cdot\boldsymbol{c}(Ref, Rec)\cdot\boldsymbol{s}(Ref, Rec)
\end{equation}
where $Rec$ is the reconstructed image, $Ref$ denotes the reference image and $N$ is the total number of image pixels. The SSIM index is a multiplicative combination of the luminance term, the contrast term, and the structural term (details shown in \cite{wang2004image}).

\subsection{The Reconstruction Performance of the Proposed SLR-Net}
\begin{figure*}[htbp]
	\centerline{\includegraphics[width=0.8\linewidth]{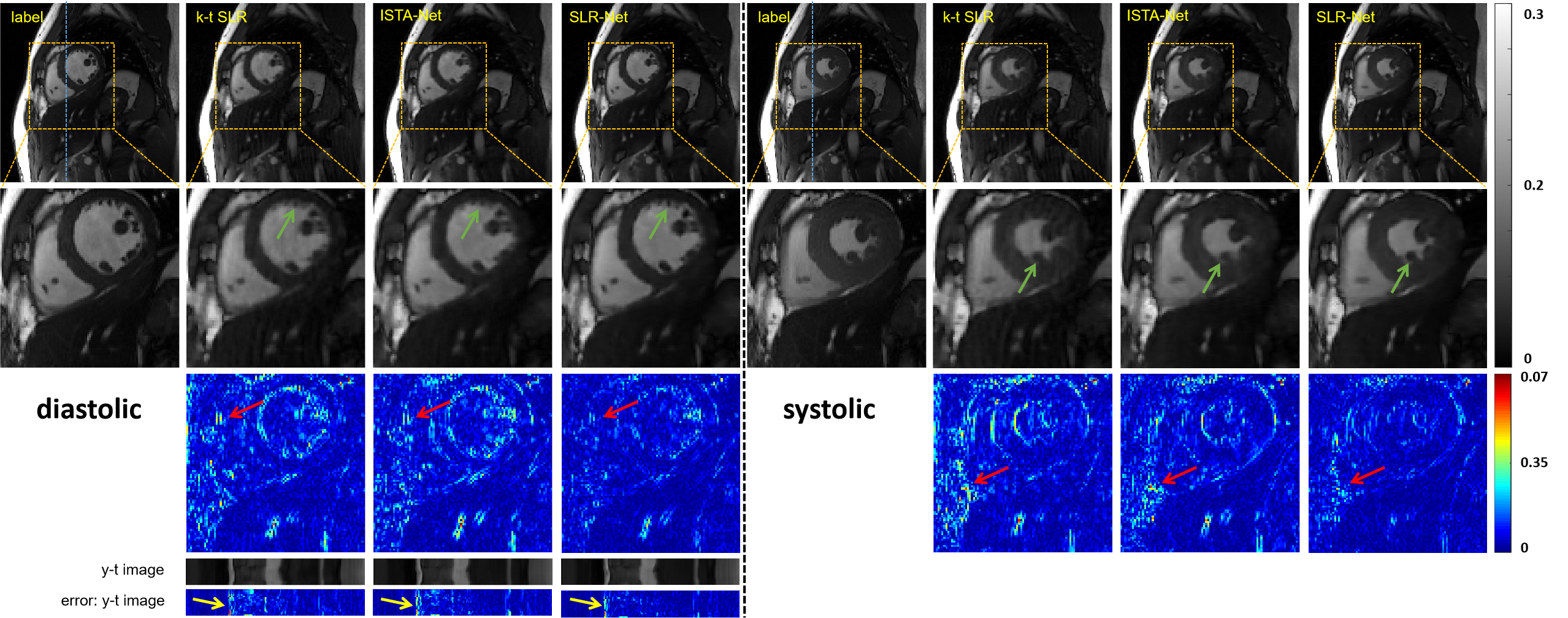}}
	\caption{The reconstruction results of the different methods (k-t SLR, ISTA-Net and the proposed SLR-Net) at 8-fold acceleration. The left half shows diastolic reconstruction, and the right half shows systolic reconstruction. The first row shows, from left to right, the ground truth and the reconstruction result of these methods. The second row shows the enlarged view of their respective heart regions framed by a yellow box. The third row shows the error map (display ranges [0, 0.07]). The y-t image (extraction of the 124th slice along the y and temporal dimensions) and the error of y-t image are also given for each signal to show the reconstruction performance in the temporal dimension. \label{results_SLR_Net}}
\end{figure*}

To demonstrate the efficacy of the proposed deep unrolling method, we compared it with a state-of-the-art CS-LR method k-t SLR \cite{lingala2011accelerated} and a sparse-based CNN method ISTA-Net \cite{Zhang_2018_CVPR}. The reconstruction results of these methods at 8-fold acceleration are shown in Fig.\ref{results_SLR_Net}. 
We present both diastolic and systolic reconstruction results to demonstrate the performance of different heart phases. The left half shows diastolic reconstruction results, and the right half shows systolic reconstruction results. The first row shows, from left to right, the ground truth and the reconstruction result of these methods. The second row shows the enlarged view of their respective heart regions framed by a yellow box. The third row shows the error map (display ranges [0, 0.07]). The y-t image (extraction of the 124th slice along the y and temporal dimensions) and the error of y-t image are also given for each signal to show the reconstruction performance in the temporal dimension. The reconstruction performance of the two deep learning based methods (ISTA-Net and SLR-Net) is better than that of the traditional iterative method (k-t SLR), which can be clearly seen from the error maps. The comparison between the two deep learning methods shows that SLR-Net is better than ISTA-Net in both detail retention and artifact removal (as shown by the green and red arrows). This indicates that the deep low-rank prior plays an important role in dynamic MR reconstruction. The y-t results also have consistent conclusions, as shown by the yellow arrows. 
\begin{table}[!t]
	\renewcommand{\arraystretch}{1.1}
	\renewcommand\tabcolsep{1.5pt}
	\caption{\textcolor{black}{The average MSE, PSNR, SSIM and Reconstruction time of k-t SLR, ISTA-Net and SLR-Net on the test dataset at 8-fold acceleration (mean$\pm$std).}}
	\label{quan_SLR_Net}
	\centering
	\textcolor{black}{\begin{tabular}{l|cccc}\hline\hline
			Methods&MSE(*e-5)&PSNR&SSIM&Time(s)\\\hline
			k-t SLR&$9.49\pm3.05$&$40.65\pm1.99$&$	0.9502\pm0.0067$&$197.49$\\
			ISTA-Net&$5.64\pm1.65$&$42.68\pm1.33$&$0.9710\pm0.0057$&$\bm{0.23}$\\
			SLR-Net&$\bm{4.12\pm1.41}$&$\bm{44.12\pm1.55}$&$\bm{0.9775\pm0.0056}$&$0.66$\\
			\hline\hline 
	\end{tabular}}
\end{table}

We also provided quantitative evaluations in Table \ref{quan_SLR_Net}. One can see that the SLR-Net achieves optimal quantitative evaluations. Although the introduction of deep low-rank increases the amount of computation, the effect on the reconstruction time is very small and can even be ignored compared with the reconstruction time of k-t SLR. In both qualitative and quantitative results, we can draw a conclusion: the first scheme of deep low-rank prior, an unrolling sparse and low-rank network, can effectively explore the low-rank prior of dynamic data, thus improving the reconstruction performance.

\subsection{The Reconstruction Performance of the Proposed Plug-and-play LR Module}
To demonstrate the efficacy of the proposed plug-and-play LR module, we embedded it in the state-of-the-art CNN-based DC-CNN \cite{schlemper2018deep} and CRNN \cite{qin2018convolutional} methods at location numbered 2 (L2), where we can obtain optimal performance as we explored in the following Section IV.A. 

\begin{figure}[htbp]
	\centerline{\includegraphics[width=1.0\linewidth]{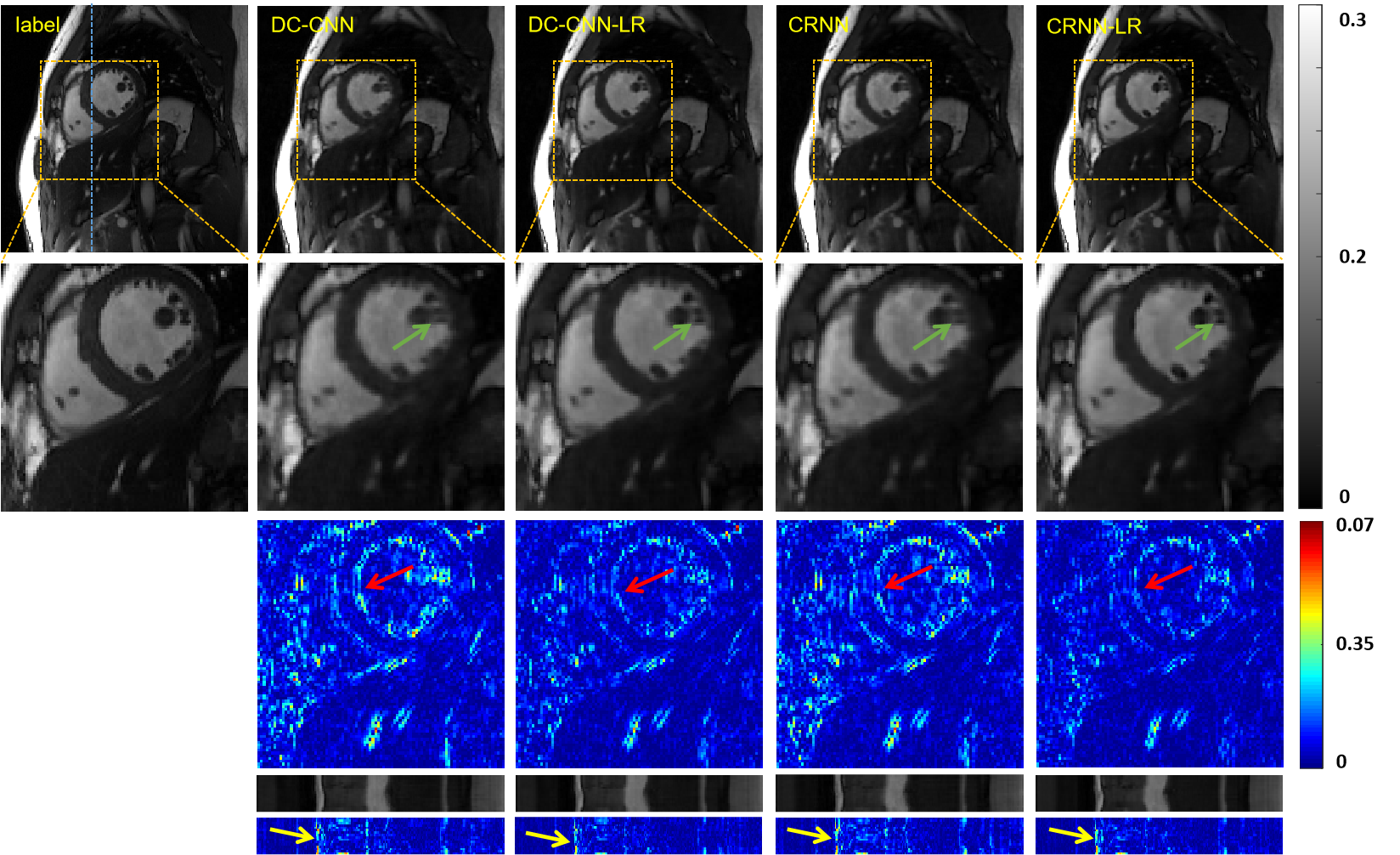}}
	\caption{The reconstruction results of the different methods (DC-CNN, DC-CNN-LR, CRNN and CRNN-LR) at 8-fold acceleration. The first row shows, from left to right, the ground truth, the reconstruction result of these methods. The second row shows the enlarged view of their respective heart regions framed by a yellow box. The third row shows the error map (display ranges [0, 0.07]). The y-t image (extraction of the 124th slice along the y and temporal dimensions) and the error of y-t image are also given for each signal to show the reconstruction performance in the temporal dimension. \label{results_LR_Module}}
\end{figure}

The reconstruction results of CNN-LR at 8-fold acceleration are shown in Fig.\ref{results_LR_Module}.
It is obvious to observe that by embedding LR module, the reconstruction results exhibit more details (as shown in the green arrow) and the smoothness is eliminated. The quantitative evaluations were also provided in Table \ref{quan_LR_Module}. All three performance metrics were significantly improved (DC-CNN: 2.0e-5 lower in MSE, 1.4dB higher in PSNR and 0.009 higher in SSIM. CRNN:  2.7e-5 lower in MSE, 2.1dB higher in PSNR and 0.01 higher in SSIM). Both qualitatively and quantitatively, the second scheme of deep low-rank prior achieves superior reconstruction performance.

DC-CNN and CRNN have consistent conclusions: The embedding of the LR module can effectively improve the reconstruction results. This lightweight LR module enables the neural networks to quickly access low-rank prior of dynamic data. It is also very easy to embed in other dynamically correlated deep learning models because it does not require changes to the topology of the network.
\begin{table}[!t]
	\renewcommand{\arraystretch}{1.1}
	\renewcommand\tabcolsep{1.5pt}
	\caption{\textcolor{black}{The average MSE, PSNR and SSIM of DC-CNN/DC-CNN-LR and CRNN/CRNN-LR on the test dataset at 8-fold acceleration (mean$\pm$std).}}
	\label{quan_LR_Module}
	\centering
	\textcolor{black}{\begin{tabular}{l|ccc}\hline\hline
			Methods&MSE(*e-5)&PSNR&SSIM\\\hline
			DC-CNN&$7.49\pm2.24$&$41.46\pm1.36$&$	0.9644\pm0.0070$\\
			DC-CNN-LR&$\bm{5.43\pm1.44}$&$\bm{42.89\pm1.42}$&$\bm{0.9736\pm0.0059}$\\\hline
			CRNN&$7.18\pm2.12$&$41.63\pm1.34$&$0.9668\pm0.0063$\\
			CRNN-LR&$\bm{4.45\pm1.43}$&$\bm{43.75\pm1.45}$&$\bm{0.9768\pm0.0054}$\\
			\hline\hline 
	\end{tabular}}
\end{table}

\section{Discussion} 
\subsection{The Locations Where the LR Module is Embedded}
\begin{figure}[htbp]
	\centerline{\includegraphics[width=1\linewidth]{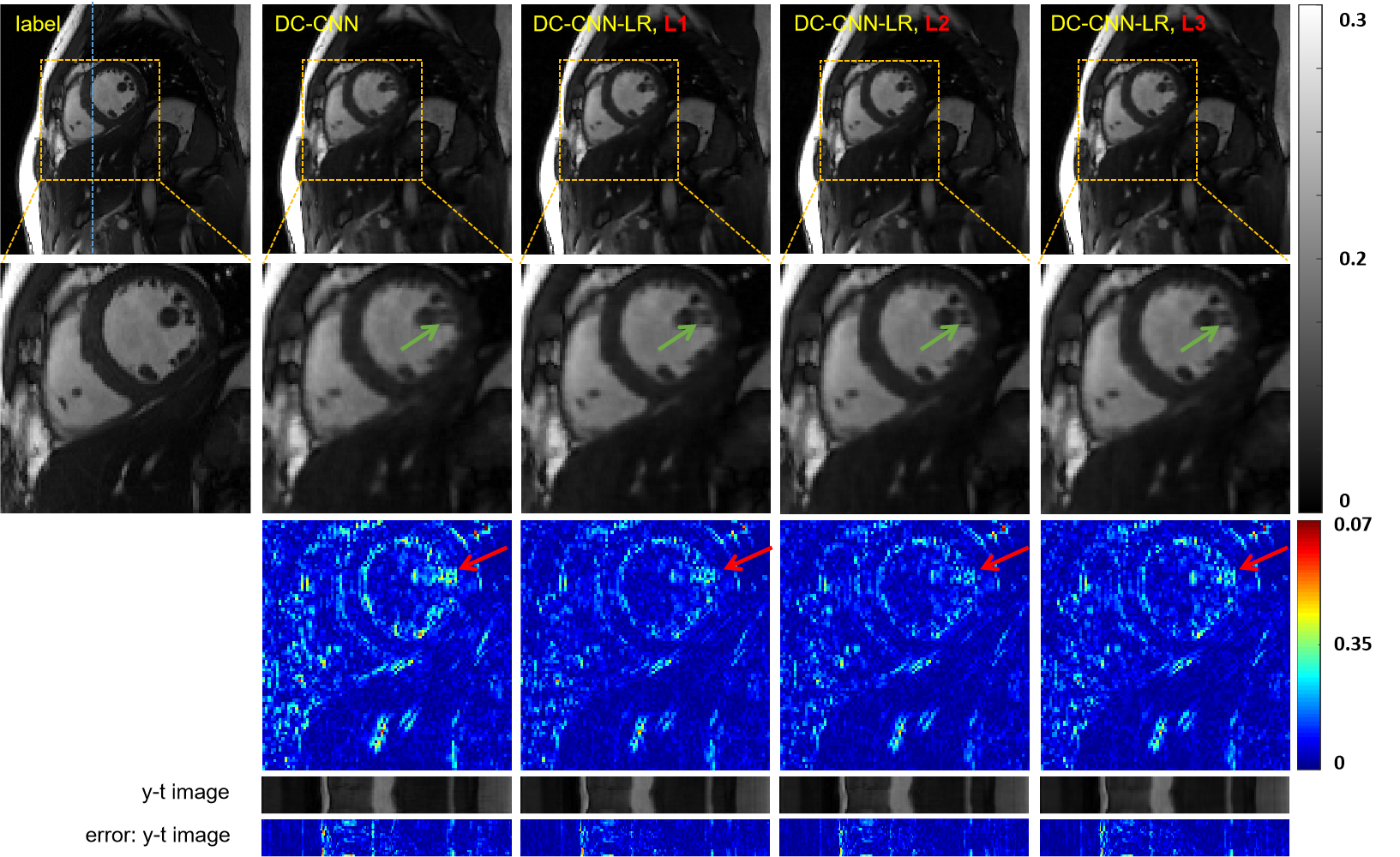}}
	\caption{The reconstruction results of DC-CNN-LR with different embedding locations of LR modules at 8-fold acceleration. The first row shows, from left to right, the ground truth, the reconstruction results of DC-CNN and DC-CNN-LR with three different locations. The second row shows the enlarged view of their respective heart regions framed by a yellow box. The third row shows the error map (display ranges [0, 0.07]). The y-t image (extraction of the 124th slice along the y and temporal dimensions) and the error of y-t image are also given for each signal to show the reconstruction performance in the temporal dimension. \label{Where_better}}
\end{figure}

In the second scheme, the LR Module can be embedded into any other dynamic deep networks. However, whether the LR module is embedded anywhere works, or where it is embedded to achieve the best reconstruction performance is still unknown. In this section, without loss of generality, we take DC-CNN as an example to explore the reconstruction performance when the LR module is embedded in different locations (L1, L2, and L3 as shown in Fig. \ref{LR-Module}). Three DC-CNN-LR models were trained under exactly the same conditions. The only difference between them lay in the different embedding locations of LR modules. As can be seen from the reconstruction results in Fig.\ref{Where_better}, no matter where LR is embedded, improved results can be achieved. This fully demonstrates the effectiveness and universality of the proposed plug-and-play LR module. The three DC-CNN-LR models achieved very similar reconstruction results, but the L2 model performed better in minimal detail, as shown by the green arrows. The quantitative evaluations are provided in Table \ref{quan_where_better}. The quantitative indicators of the three LR models are all consistently superior to the original DC-CNN model. The internal comparison of the three LR models shows that their quantitative indicators are very similar. The L2 model achieves the best quantization performance with a slight advantage. The quantization performance of the L3 model is slightly weak, which may be because the LR module is embedded behind the DC module, which slightly damages the performance of the DC module. 

Based on the above results, the following recommendation is made for embedding LR modules: Embedding the LR module in a network can improve the reconstruction performance, but try to avoid embedding behind the DC module, which might counteract the role of DC module. 
\begin{table}[!t]
	\renewcommand{\arraystretch}{1.1}
	\renewcommand\tabcolsep{1.5pt}
	\caption{\textcolor{black}{The average MSE, PSNR and SSIM of DC-CNNand DC-CNN-LR (L1, L2, L3) on the test dataset at 8-fold acceleration (mean$\pm$std).}}
	\label{quan_where_better}
	\centering
	\textcolor{black}{\begin{tabular}{l|ccc}\hline\hline
			Methods&MSE(*e-5)&PSNR&SSIM\\\hline
			DC-CNN&$7.49\pm2.24$&$41.46\pm1.36$&$	0.9644\pm0.0070$\\
			DC-CNN-LR, L1&$5.40\pm1.43$&$42.88\pm1.45$&$0.9725\pm0.0054$\\
			DC-CNN-LR, L2&$\bm{5.36\pm1.44}$&$\bm{42.89\pm1.42}$&$\bm{0.9736\pm0.0059}$\\
			DC-CNN-LR, L3&$5.96\pm1.75$&$42.43\pm1.31$&$0.9701\pm0.0058$\\
			\hline\hline 
	\end{tabular}}
\end{table}

\subsection{Which of the Two Schemes Performs Better}
\begin{figure}[htbp]
	\centerline{\includegraphics[width=0.8\linewidth]{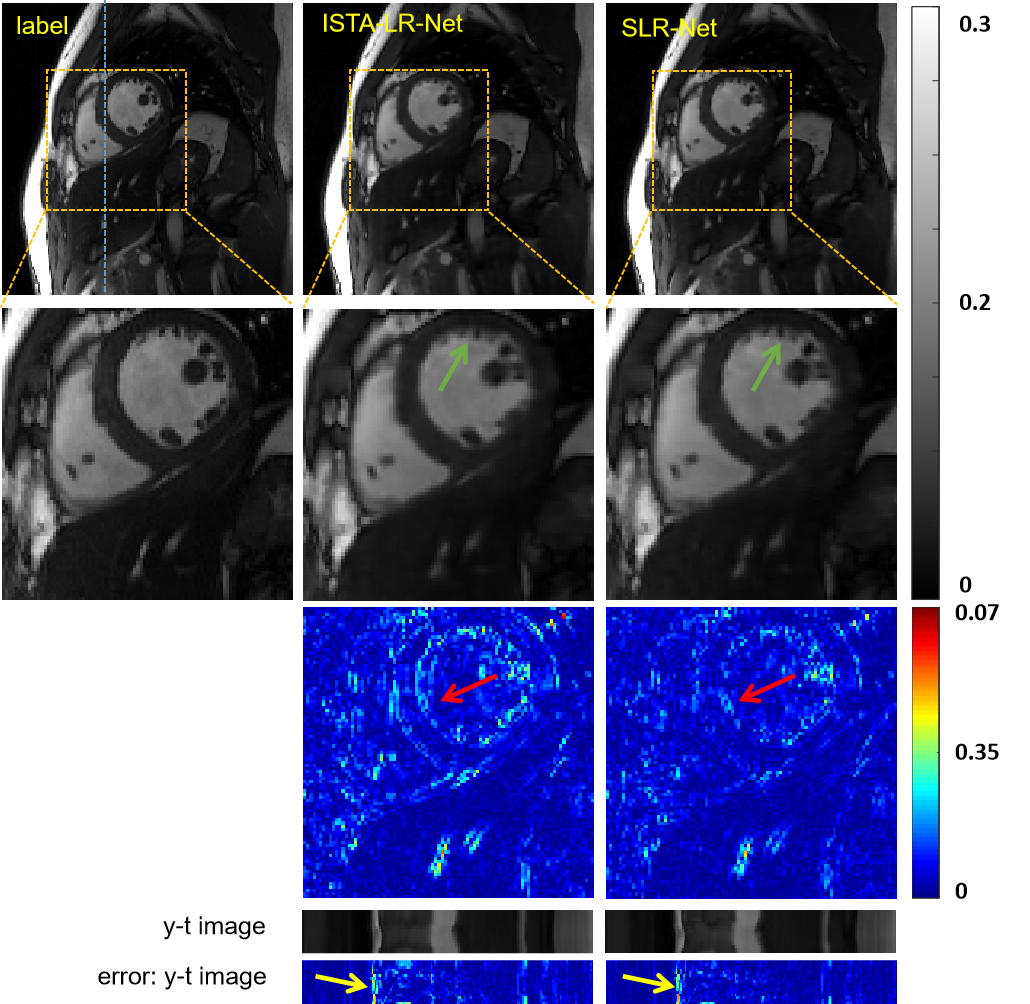}}
	\caption{The reconstruction results of ISTA-LR-Net and SLR-Net at 8-fold acceleration. The first row shows, from left to right, the ground truth, the reconstruction result of these methods. The second row shows the enlarged view of their respective heart regions framed by a yellow box. The third row shows the error map (display ranges [0, 0.07]). The y-t image (extraction of the 124th slice along the y and temporal dimensions) and the error of y-t image are also given for each signal to show the reconstruction performance in the temporal dimension. \label{Which_better}}
\end{figure}

The above sections demonstrate that both schemes that introduce deep low-rank prior can lead to improvements in the reconstructed results. In this section, we will explore which of these two schemes works better. To be fair, the same baseline is needed, and we regarded ISTA-Net as the baseline method. Obviously, SLR-Net is the unrolling version of deep low-rank prior based on ISTA-Net. Its plug-and-play version of deep low-rank prior is ISTA-LR-Net as shown in Fig.\ref{LR-Module} (d). The reconstruction results of these two models at 8-fold acceleration are shown in Fig.\ref{Which_better}. Both models have achieved better reconstruction performance, but SLR-Net has some advantages in preserving detail, as shown by the green arrow. SLR-Net also has certain progress in quantitative indicators, as shown in Table \ref{quan_which_better}. Therefore, we come to the conclusion that under the same baseline, unrolled deep low-rank prior is superior to plug-and-play deep low-rank prior. We conjecture that the main reason is that in the plug-and-play scheme, a first-order Taylor expansion (Eq.\ref{ite2}) is used to approximate the regularizer $\|Dx\|_1$, which is a relaxed approximation due to its nonsmoothness, while in the unrolling scheme, the regularizer is exactly used (Eq.\ref{eq_13}).

Both schemes have their own advantages and applicable scenarios: Unrolled deep low-rank prior requires a highly customized optimization model and solution algorithm. If all of these are available, we recommend using the first schemes to get the best reconstruction results. The highlight of plug-and-play deep low-rank prior is its portability. If it is non-trivial to build an unrolled SLR network in other imaging systems, plug-and-play deep low-rank prior is recommended to utilize the low-rank prior quickly.
\begin{table}[!t]
	\renewcommand{\arraystretch}{1.1}
	\renewcommand\tabcolsep{1.5pt}
	\caption{\textcolor{black}{The average MSE, PSNR and SSIM of ISTA-LR-Net and SLR-Net on the test dataset at 8-fold acceleration (mean$\pm$std).}}
	\label{quan_which_better}
	\centering
	\textcolor{black}{\begin{tabular}{l|ccc}\hline\hline
			Methods&MSE(*e-5)&PSNR&SSIM\\\hline
			ISTA-LR-Net&$4.54\pm1.46$&$43.66\pm1.45$&$	0.9758\pm0.0055$\\
			SLR-Net&$\bm{4.12\pm1.41}$&$\bm{44.12\pm1.55}$&$\bm{0.9775\pm0.0056}$\\
			\hline\hline 
	\end{tabular}}
\end{table}

\subsection{Higher Acceleration: 10-fold and 12-fold} 
\begin{figure*}[htbp]
	\centerline{\includegraphics[width=0.8\linewidth]{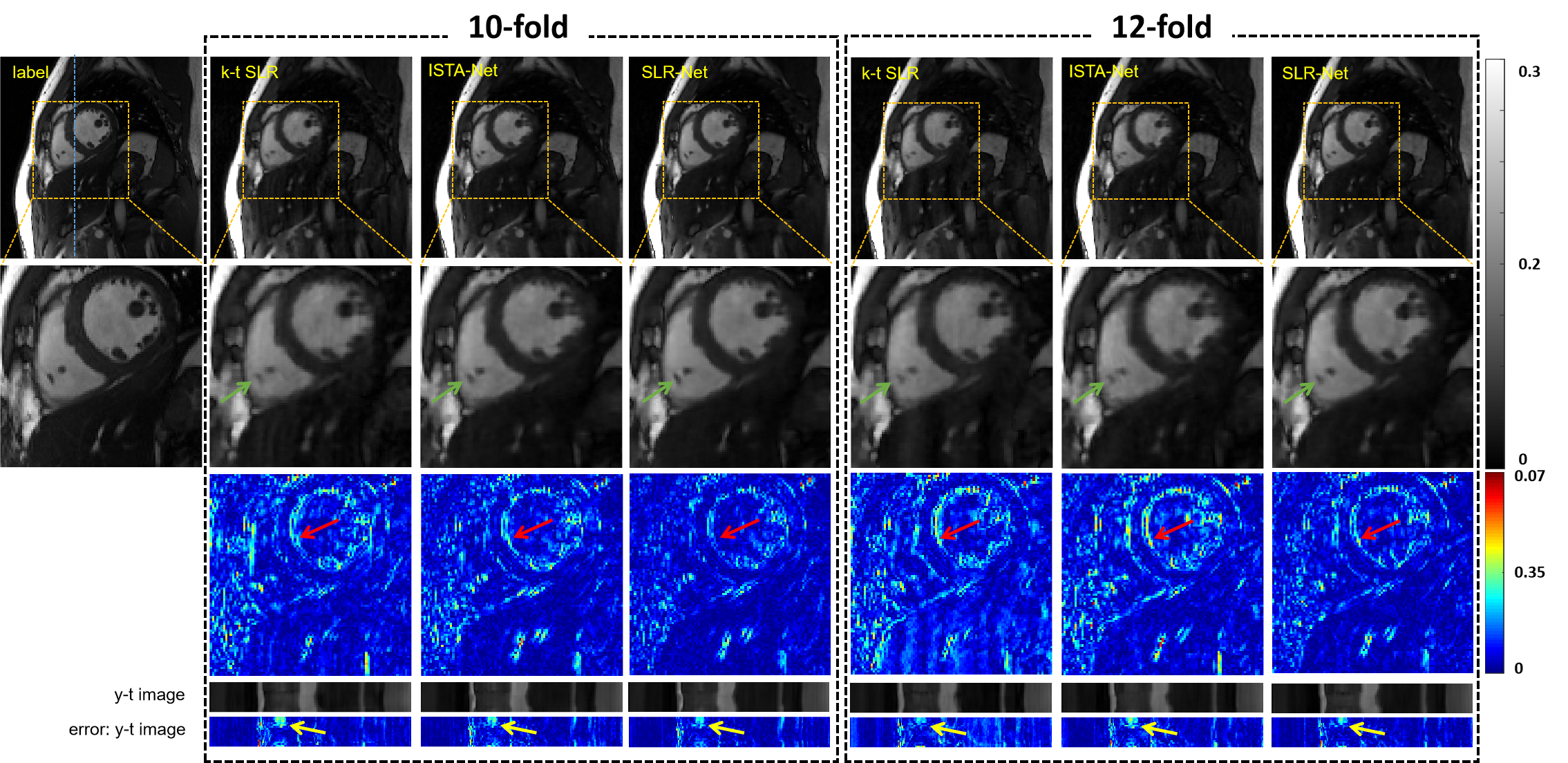}}
	\caption{The reconstruction results of the proposed SLR-Net at 10-fold and 12-fold acceleration. The left half shows 10-fold reconstruction, and the right half shows 12-fold reconstruction. The first row shows, from left to right, the ground truth and the reconstruction result of these methods. The second row shows the enlarged view of their respective heart regions framed by a yellow box. The third row shows the error map (display ranges [0, 0.07]). The y-t image (extraction of the 124th slice along the y and temporal dimensions) and the error of y-t image are also given for each signal to show the reconstruction performance in the temporal dimension. \label{higher_acc}}
\end{figure*}

The proposed deep low-rank prior methods can make use of both sparse and low rank prior of dynamic MR data, which not only improves the reconstruction performance, but also increases the acceleration, because more regularization terms are introduced into the optimization problem. Without loss of generality, we explore the reconstruction performance at higher accelerations on the proposed SLR-Net. The 10-fold and 12-fold accelerated reconstruction results can be found in Fig. \ref{higher_acc}. Our proposed SLR-Net still achieves superior reconstruction performance, whether it is 10-fold or 12-fold. At 10-fold acceleration, the proposed SLR-Net can achieve satisfactory reconstruction results. The anatomical details of the heart tissue can be easily found, and the blurring is not serious. At 12-fold acceleration, the proposed SLR-Net can still achieve acceptable reconstruction results, although it is a little vague, most of the details are well preserved. The quantitative indicators are provided in Table \ref{quan_higher_acc}, from which we can see that our proposed SLR-Net still achieves excellent quantitative performance at higher accelerations. Although we did not show the results of higher accelerations under the plug-and-play scheme, similar conclusions can be drawn in this scheme.
\begin{table}[!t]
	\renewcommand{\arraystretch}{1.1}
	\renewcommand\tabcolsep{1.5pt}
	\caption{\textcolor{black}{The average MSE, PSNR and SSIM of SLR-Net at 10-fold and 12-fold acceleration on the test dataset (mean$\pm$std).}}
	\label{quan_higher_acc}
	\centering
	\textcolor{black}{\begin{tabular}{ll|ccc}\hline\hline
			&Methods&MSE(*e-5)&PSNR&SSIM\\\hline
			&k-t SLR&$11.13\pm3.86$&$39.73\pm1.59$&$	0.9422\pm0.0162$\\
			10x$\ \ $&ISTA-Net&$6.98\pm2.11$&$41.77\pm1.38$&$0.9664\pm0.0069$\\
			&SLR-Net&$\bm{5.71\pm1.77}$&$\bm{42.66\pm1.42}$&$\bm{0.9712\pm0.0063}$\\\hline
			&k-t SLR&$25.69\pm8.62$&$36.19\pm1.69$&$	0.8563\pm0.0428$\\
			12x$\ \ $&S-Net&$9.92\pm2.40$&$40.21\pm1.41$&$0.9524\pm0.0073$\\
			&SLR-Net&$\bm{7.64\pm2.31}$&$\bm{41.38\pm1.37}$&$\bm{0.9624\pm0.0071}$\\
			\hline\hline 
	\end{tabular}}
\end{table}

\subsection{The limitations of the Proposed Deep Low-rank Prior} 
Although the two proposed deep low-rank prior methods achieve improved reconstruction results, they still have the following limitations: 1) The low-rank module greatly occupies GPU memory, which puts forward higher requirements for hardware. 2) This work focused on single-channel dynamic MR imaging. It leads to the underutilization of spatial variance coil sensitivity provided by the phased array coil, which has been proven to play an important role in fast MRI. Nevertheless, it is difficult to explore low-rank prior and parallel image simultaneously under the current hardware conditions because both require a lot of GPU memory. 3) In the proposed LR module, we selected learned top k singular values in the singular vector. However, there are many other SVT strategies can be chosen. Unfortunately, we did not explore which SVT strategy is optimal. In future work, these limitations will be addressed.

\section{Conclusion and Outlook}
In this paper, we explore deep low-rank prior in dynamic MR imaging to obtain improved reconstruction results. In particular, we propose a learned low-rank prior with two novel and distinct schemes to introduce it into deep network architectures in an unrolling manner and a plug-and-play manner. In the unrolling manner, we propose a model-based unrolling sparse and low-rank network for dynamic MR imaging, dubbed SLR-Net. The SLR-Net is defined over a deep network flow graph, which is unrolled from the iterative procedures in the Iterative Shrinkage-Thresholding Algorithm (ISTA) for optimizing a sparse and low-rank based dynamic MRI model. In the plug-and-play manner, we propose a plug-and-play LR network module that can be easily embedded into any other dynamic MR neural networks without changing the neural network paradigm.  Experimental results show that both schemes can improve the reconstruction results qualitatively and quantitatively. To the best of our knowledge, this is the first time that a deep low-rank prior has been applied in dynamic MR imaging.

\ifCLASSOPTIONcaptionsoff
  \newpage
\fi



\bibliographystyle{IEEEtran}
\bibliography{IEEEabrv,SLR_Net}
\end{document}